\documentclass[twocolumn,showpacs,preprintnumbers,amsmath,amssymb,superscriptaddress]{revtex4}

\usepackage{graphicx}
\usepackage{dcolumn}
\usepackage{bm}
\begin{document}

\title{Quantum Dynamics of a Microwave Driven Superconducting Phase Qubit
Coupled to a Two-Level System}
\author{Guozhu Sun}
\email{gzsun@nju.edu.cn}
\affiliation{Research Institute of Superconductor Electronics,School of Electronic
Science and Engineering, Nanjing University, Nanjing 210093, China}
\affiliation{Department of Physics and Astronomy, University of Kansas, Lawrence, KS
66045, USA}
\author{Xueda Wen}
\affiliation{National Laboratory of Solid State Microstructures, School of Physics,
Nanjing University, Nanjing 210093, China}
\author{Bo Mao}
\affiliation{Department of Physics and Astronomy, University of Kansas, Lawrence, KS
66045, USA}
\author{Zhongyuan Zhou}
\affiliation{Department of Chemistry, University of Kansas, Lawrence, KS 66045, USA}
\author{Yang Yu}
\affiliation{National Laboratory of Solid State Microstructures, School of Physics,
Nanjing University, Nanjing 210093, China}
\author{Peiheng Wu}
\affiliation{Research Institute of Superconductor Electronics,School of Electronic
Science and Engineering, Nanjing University, Nanjing 210093, China}
\author{Siyuan Han}
\email{han@ku.edu}
\affiliation{Department of Physics and Astronomy, University of Kansas, Lawrence, KS
66045, USA}
\date{\today}

\begin{abstract}
We present an analytical and comprehensive description of the quantum
dynamics of a microwave resonantly driven superconducting phase qubit
coupled to a microscopic two-level system (TLS), covering a wide range of
the external microwave field strength. Our model predicts several interesting
phenomena in such an ac driven four-level bipartite system including anomalous
Rabi oscillations, high-contrast beatings of Rabi oscillations, and
extraordinary two-photon transitions. Our experimental results
in a coupled qubit-TLS system agree quantitatively very well with the predictions of the theoretical model.
\end{abstract}

\maketitle

Microscopic two-level systems (TLSs) originating from variety of disorders
are ubiquitous in solid state devices. In particular, the TLS plays a
double-faced character in the implementation of quantum information
processor based on the superconducting qubits \cite{RevModPhys.73.357,PhysicsToday.58.11,Nature.453.1031}.
On one side, the TLSs
are harmful for the coherent control of qubit states \cite{PhysRevLett.93.077003,PhysRevLett.95.210503,PhysRevB.78.144506,PhysRevLett.95.127002, PhysRevLett.96.047001,PhysRevB.79.094520,muller:134517}%
. Although significant progress has been made to reduce the TLSs in the
superconducting qubits \cite{QuantumInfProcess.8.81,QuantumInfProcess.8.117}
it seems however difficult to completely eliminate them in
the foreseeable future. On the other side, the TLSs can play a useful
role because of their relatively long coherence time \cite%
{PhysRevLett.97.077001,NaturePhysics.4.523,PhysRevLett.101.157001,Nat.Commun.,PhysRevB.80.094507}%
. It has been demonstrated that the TLSs in the tunnel barrier of a
Josephson junction can function as naturally formed qubits \cite%
{PhysRevLett.97.077001}, quantum memory cells \cite%
{NaturePhysics.4.523}, and quantum beam splitters \cite{Nat.Commun.}. In
either case a thorough understanding of the quantum dynamics of the coupled
qubit-TLS system is critical. What's more, the quantum dynamics of a
resonantly driven coupled quantum bipartite system is important not only to
fundamental physics \cite{Nielsen,NaturePhysics.4.686} but also to
applications such as building scalable quantum information
processors based on qubits \cite%
{RevModPhys.73.357,PhysicsToday.58.11,Nature.453.1031,NaturePhysics.6.13}.
However, previous theoretical studies mainly focused on a narrow
range of the ac driving strengths and only numerical simulations were
performed, leaving the physics picture unclear. Consequently, the
experimental investigations were incomplete although interesting two-photon
Rabi oscillations and anomalous Rabi oscillations have been observed \cite%
{PhysRevLett.93.077003,PhysRevB.80.172506,PhysRevB.81.100511}. Here,
we present an analytical and comprehensive description of the quantum
dynamics of a resonantly driven qubit-TLS system that covers a wide range of
ac field and qubit-TLS coupling strengths. More importantly, we
have experimentally investigated a superconducting phase qubit coupled to the
TLSs with all critical system parameters calibrated independently and the
data strongly support the results of our theoretical analysis. Therefore,
the results reported here, both theoretical and experimental, provide many
insights into the dynamics of resonantly driven coupled four-level bipartite
quantum systems.

We start by modeling the qubit-TLS system with the Hamiltonian \cite%
{PhysRevLett.101.157001,PhysRevB.80.172506,PhysRevB.81.100511,
EPL.71.21,PhysRevB.72.024526,New.J.Phys.8.103,PhysRevB.80.094507} $%
H(t)=H_{q}(t)+H_{T}+H_{q-T}$. The qubit's Hamiltonian is $H_{q}(t)=-\frac{%
\hbar }{2}\omega _{q}\sigma _{z}^{q}+\hbar \Omega _{m}\cos \omega t\sigma
_{x}^{q}$, where $\hbar $ is Planck's constant divided by $2\pi $, $\hbar
\omega _{q}$ is the energy level spacing of the qubit, $\Omega _{m}/2\pi $
is the microwave induced Rabi frequency, and $\omega $ is the microwave
frequency. The Hamiltonian of the TLS can be written as $H_{T}=-\frac{\hbar
}{2}\omega _{T}\sigma _{z}^{T}$, where $\hbar \omega _{T}$ is the energy
level spacing of the TLS. The interaction Hamiltonian then is $H_{q-T}=\hbar
\Omega _{c}\sigma _{x}^{q}\otimes \sigma _{x}^{T}$, where $\Omega _{c}$ is
the coupling strength between the qubit and the TLS, $\sigma _{x,y,z}^{q}$ ($%
\sigma _{x,y,z}^{T}$) are the Pauli operators acting on the states of the
qubit (TLS). Such $\sigma _{x}$ coupling is usually found in NMR and other
systems \cite{RevModPhys.73.357,RevModPhys.76.1037}. To understand the
underlying physics more clearly, we choose the interaction picture and make
a transformation to a rotating frame, denoting the ground state and excited
state of the qubit (TLS) as $|0\rangle $ and $|1\rangle $ ($|g\rangle $ and $%
|e\rangle $), respectively. In the resonant case, i.e., $\omega =\omega
_{q}=\omega _{T}$, the Hamiltonian for the coupled system can be simplified
as \cite{PhysRevB.80.094507}
\begin{equation}
H^{\prime }=\hbar \left(
\begin{array}{cccc}
0 & \Omega _{m}/2 & 0 & 0 \\
\Omega _{m}/2 & 0 & \Omega _{c} & 0 \\
0 & \Omega _{c} & 0 & \Omega _{m}/2 \\
0 & 0 & \Omega _{m}/2 & 0 \\
\end{array}%
\right) ,  \label{simple form}
\end{equation}%
where the basis states are $|0g\rangle $, $|1g\rangle $, $|0e\rangle $, and $%
|1e\rangle $ (inset of Fig. 1(a)). It is noticed that the Hamiltonian (\ref%
{simple form}) has the same form as that of the four coupled quantum
pendulums \cite{Feynman}. The simplest way to analyze such system is to find
the stationary solutions without considering dissipation, which
only affects the amplitude of the probability. The eigenvalues of
the Hamiltonian (\ref{simple form}) can be easily obtained in the form: $%
\lambda _{1}=(\Omega _{mc}+\Omega _{c})/2$, $\lambda _{2}=(\Omega
_{mc}-\Omega _{c})/2$, $\lambda _{3}=-(\Omega _{mc}-\Omega _{c})/2$, $%
\lambda _{4}=-(\Omega _{mc}+\Omega _{c})/2$, with $\Omega _{mc}=\sqrt{\Omega
_{m}^{2}+\Omega _{c}^{2}}$. The time evolution of the system can be
described by $|\Psi (t)\rangle =\sum_{i=1}^{4}c_{i}|\psi _{i}\rangle
e^{-i\lambda _{i}t}$, where $|\psi (i)\rangle $ is the eigenfunction of the
Hamiltonian (1) corresponding to $\lambda _{i}$. Thus we can obtain the
probability of being in state $|\phi \rangle $ ($\phi =0$g, 1g, 0e, 1e):
\begin{equation}
P_{\phi }=\sum_{i,j}c_{j}^{\ast }c_{i}\langle \psi _{j}|\phi \rangle \langle
\phi |\psi _{i}\rangle e^{-i(\lambda _{i}-\lambda _{j})t}.
\end{equation}%
Note that the temporal oscillation of $P_{\phi }$ is composed of $%
C_{4}^{2}=6$ frequencies. However, only four frequencies are observable
because there are two pairs of double degeneracies: $\Omega
_{1}=|\lambda _{1}-\lambda _{4}|=\Omega _{mc}+\Omega _{c}$, $\Omega
_{2}=|\lambda _{2}-\lambda _{3}|=\Omega _{mc}-\Omega _{c}$, $\Omega
_{3}=|\lambda _{1}-\lambda _{3}|=|\lambda _{2}-\lambda _{4}|=\Omega _{mc}$,
and $\Omega _{4}=|\lambda _{1}-\lambda _{2}|=|\lambda _{3}-\lambda
_{4}|=\Omega _{c}$. Assuming the system is initially prepared in $%
|\Psi (0)\rangle =|0g\rangle $, it is easy to obtain $P_{\phi }$: {\small
\begin{equation}
\left\{
\begin{array}{ll}
P_{0g}=\frac{4a^{2}b^{2}}{N^{2}}[\frac{a^{4}+b^{4}}{2a^{2}b^{2}}+\frac{a^{2}%
}{2b^{2}}\cos \Omega _{1}t+\frac{b^{2}}{2a^{2}}\cos \Omega _{2}t+\cos \Omega
_{3}t &  \\
+\cos \Omega _{4}t] &  \\
P_{1g}=\frac{4a^{2}b^{2}}{N^{2}}[1-\frac{1}{2}\cos \Omega _{1}t-\frac{1}{2}%
\cos \Omega _{2}t-\cos \Omega _{3}t+\cos \Omega _{4}t] &  \\
P_{0e}=\frac{4a^{2}b^{2}}{N^{2}}[1+\frac{1}{2}\cos \Omega _{1}t+\frac{1}{2}%
\cos \Omega _{2}t-\cos \Omega _{3}t-\cos \Omega _{4}t] &  \\
P_{1e}=\frac{4a^{2}b^{2}}{N^{2}}[\frac{a^{4}+b^{4}}{2a^{2}b^{2}}-\frac{a^{2}%
}{2b^{2}}\cos \Omega _{1}t-\frac{b^{2}}{2a^{2}}\cos \Omega _{2}t+\cos \Omega
_{3}t &  \\
-\cos \Omega _{4}t] &  \\
\end{array}%
\right. ,  \label{pulse}
\end{equation}%
} where $a=\Omega _{m}/2$, $b=(\Omega _{mc}+\Omega _{c})/2$ and $N=\left[
\Omega _{m}^{2}+(\Omega _{mc}+\Omega _{c})^{2}\right] /2$ being the
normalization factor. Apparently, the populations will oscillate in time
with four frequencies. These analytical results form the foundation for a
thorough understanding of the coupled qubit-TLS system.
Moreover, these general results are also valid for a wide variety of four-level
quantum bipartite systems with resonant ac drive. Based on the results many
seemingly counterintuitive phenomena in this type of four-level systems become
straightforward to understand both qualitatively and quantitatively.

In the experiments involving the TLSs, the quantity one usually
measures is the probability of finding the qubit in state $%
|1\rangle $, i.e., $P_{1}=P_{1g}+P_{1e}$: 
\begin{equation}
\begin{array}{ll}
P_{1}=\frac{4a^{2}b^{2}}{N^{2}}\left[ \frac{(a^{2}+b^{2})^{2}}{2a^{2}b^{2}}-%
\frac{a^{2}+b^{2}}{2b^{2}}\cos \Omega _{1}t-\frac{a^{2}+b^{2}}{2a^{2}}\cos
\Omega _{2}t\right] . &
\end{array}%
\end{equation}%
%
%
It is interesting to notice that although there are four frequency
components in each $P_{\phi }$ (Fig. 2(a) and Fig. 2(b)), only two of them
appear in $P_{1}$ (Fig. 2(d)). Moreover, $\Omega _{m}$ and $\Omega _{c}$
define three regimes for the bipartite system, which show significantly
different behaviors: (i) The strong field regime, $\Omega _{m}/\Omega
_{c}\gg 1$; (ii) The intermediate field regime \cite%
{PhysRevLett.93.077003,PhysRevB.81.100511}, $\Omega _{m}/\Omega _{c}\approx
1 $ ; (iii) The weak field regime \cite{PhysRevB.80.172506}, $\Omega
_{m}/\Omega _{c}\ll 1$. It should be pointed out that previous works
focused only on one of the regimes thereby they could not capture
the complete picture. Below, based on Eq. (4), we discuss the behaviors of
the resonantly driven qubit-TLS system in the three different regimes one by
one. The predicted phenomena are demonstrated experimentally by measuring
the spectroscopy and the coherent oscillations in a superconducting phase
qubit coupled to the TLSs. The detailed experimental setup and procedures have
been described elsewhere \cite{Nat.Commun.}. Shown in Fig. 1 are examples of
the spectroscopy and the coherent oscillations. The splitting caused by the
qubit-TLS interaction is clearly observed at 16.572 GHz
giving the coupling strength $\Omega _{c}/2\pi \approx 26.5$ MHz. Away from
the splitting Rabi oscillation induced by the microwave field has the usual
damped sinusoidal form (Fig. 1(b)). When the qubit is biased at the
splitting, anomalous oscillations are observed, with interesting features
determined by the microwave amplitude as described below.

(i)\emph{\ Strong field regime: Rabi beating }

In the strong field limit, i.e., $\Omega _{m}/\Omega _{c}\gg 1$, $b\approx
a=\Omega _{m}/2$, $P_{1}$ has a simple form:
\begin{equation}
P_{1}=\frac{1}{4}[2-\cos (\Omega _{m}+\Omega _{c})t-\cos (\Omega _{m}-\Omega
_{c})t].  \label{beating}
\end{equation}%
It is known in acoustics that beating happens as an interference between two
waves of slightly different frequencies. Here the two frequencies in Eq. ({%
\ref{beating}}) are close to each other, i.e, $\Omega _{m}+\Omega
_{c}\approx \Omega _{m}-\Omega _{c}\approx \Omega _{m}$, which satisfies the
condition of beating very well. To be more clear, we further write Eq. (\ref%
{beating}) as $P_{1}=\frac{1}{2}(1-\cos \Omega _{m}t\cos \Omega _{c}t)$.
This is demonstrated experimentally in Fig. 1(c), in which $P_{1}$
appears to oscillate at $\Omega _{m}$ with the amplitude modulated by a
much lower frequency $\Omega _{c}$.

To quantitatively characterize the Rabi beating described above, we
define $Q_{b}=\Omega _{m}/2\Omega _{c}$ as the frequency contrast
of the beating, which is the number of Rabi oscillation periods
between the two nodes of the slow varying envelope. Shown in Fig. 2(c) are
the measured oscillations of $P_{1}$ at various microwave powers.
As the microwave power is increased, $Q_{b}$ increases from about 2 to 7.
Therefore the beating becomes increasingly clear. The largest $Q_{b}
$ is determined by $\Omega _{m},$ $\Omega _{c},$ and the decoherence time.
The exponential decay of the beating envelop is due to the energy
relaxation. In the previous experiments \cite%
{PhysRevLett.93.077003,PhysRevB.81.100511,Nature.421.823}, anomalous Rabi
oscillations due to the coupling between two qubits and qubit-TLS were
reported. However, since the condition $\Omega _{m}\gg \Omega _{c}$ is not
fulfilled, $Q_{b}$ is generally less than 2 and thus no clear pattern of
Rabi beating was observed, although several theoretical works \cite%
{PhysRevB.72.024526,EPL.71.21,New.J.Phys.8.103,claudon:184503} have
predicted its existence in the superconducting qubits. Nevertheless, one can
always apply the Fourier transform (FT) to obtain the two frequency
components, as shown in Fig. 2(d). With the help of FT, low-$Q_{b}$
Rabi beatings can be revealed. Therefore, we argue that our
experiment is the first to clearly demonstrate Rabi beating in a
superconducting phase qubit. Quantum beating is usually found among the
three-level atomic systems and has been applied to resolve the
detailed internal structure of matter \cite{RevModPhys.74.1153}. Rabi beating in
the coupled qubit-TLS systems thus can be utilized as a powerful
tool in investigating the origin and properties of the TLS.

In Fig. 2(d), we find that the difference between the two Rabi
frequencies obtained from FT is exactly $2\Omega _{c}$, which is independent of the microwave power as expected from Eq. ({4}). In addition,
it is interesting to notice that in the strong field limit the population of
finding the TLS in the state $|e\rangle $ has a simple form $%
P_{e}=P_{0e}+P_{1e}=(1-\cos \Omega _{c}t)/2$, which can be viewed as Rabi
oscillation between the subspaces $\{|0g\rangle ,|1g\rangle \}$ and $%
\{|0e\rangle ,|1e\rangle \}$ \cite{PhysRevB.80.094507}. The oscillation
frequency ($\Omega _{c}$) between these two subspaces is half of that ($%
2\Omega _{c}$) between $|1g\rangle $ and $|0e\rangle $ since the probability
of finding the system in $|1g\rangle $ ($|0e\rangle $) in each subspace is
exactly $1/2$.

(ii)\emph{\ Intermediate field regime: anomalous Rabi oscillation}

In this regime, $\Omega _{m}\approx \Omega _{c}$, the two frequencies in $%
P_{1}$ are well separated. The frequency contrast of the beating $Q_{b}$ is
close to unity and no clear pattern of Rabi beating can be observed, as shown in Fig. 2(c). In particular, in the region of weak ac
driving interplay between the two frequencies results in anomalous Rabi
oscillations which have been observed previously in various
superconducting qubits \cite%
{PhysRevLett.93.077003,PhysRevB.81.100511,Nature.421.823}. We can extract $%
\Omega _{c}$ and $\Omega _{m}$ by using FT of the anomalous Rabi
oscillations. In this regime, the qubit is most strongly affected by the
TLSs. Thus, much care has to be taken when performing quantum
information processing if the Josephson junction is the intended qubit only.
Since $b$ is always greater than $a,$ the weight of the $\Omega _{1}$
component is smaller than that of $\Omega _{2}$ from Eq. (4) (indicated by
the color in Fig. 2). With $\Omega _{m}$ further decreasing to the weak
field regime, the frequency $\Omega _{1}$ disappears, and $P_{1}$ oscillates
with a single frequency $\Omega _{2}$, which will be discussed in detail
below.

(iii)\emph{\ Weak field regime\emph{:} extraordinary two-photon transitions}

In this case, $\Omega _{m}\ll \Omega _{c}$, $P_{0e}+P_{1g}=\frac{8a^{2}b^{2}%
}{N^{2}}(1-\cos \Omega _{3}t)\ll 1,$ the populations in the states $%
|1g\rangle $ and $|0e\rangle $ are very small. The system mainly evolves in
the subspace spanned by $|0g\rangle $ and $|1e\rangle $. This can be
seen more clearly by checking the eigen-wavefunction of the Hamiltonian (%
\ref{simple form}) directly. In the weak field limit, $a\ll b$, we have

\begin{equation}
\left(
\begin{array}{cccc}
|\psi _{1}\rangle \\
|\psi _{2}\rangle \\
|\psi _{3}\rangle \\
|\psi _{4}\rangle \\
\end{array}%
\right) =\frac{1}{\sqrt{2}}\left(
\begin{array}{cccc}
0 & 1 & -1 & 0 \\
1 & 0 & 0 & 1 \\
-1 & 0 & 0 & 1 \\
0 & 1 & 1 & 0 \\
\end{array}%
\right) \left(
\begin{array}{cccc}
|0g\rangle\\
|1g\rangle\\
|0e\rangle\\
|1e\rangle\\
\end{array}%
\right) .
\end{equation}%
Obviously, starting from the initial state $|0g\rangle $
the system will evolve in the subspace spanned by $%
|0g\rangle $ and $|1e\rangle $ in the form of $|\Psi (t)\rangle
=(|\psi _{2}\rangle e^{-i\lambda _{2}t}-|\psi _{3}\rangle e^{-i\lambda
_{3}t})/\sqrt{2}=\left[ (e^{-i\lambda _{2}t}+e^{-i\lambda _{3}t})|0g\rangle
+(e^{-i\lambda _{2}t}-e^{-i\lambda _{3}t})|1e\rangle \right] /2$ and it is
easy to obtain
\begin{equation}
P_{1}\approx P_{1e}=\frac{1}{2}(1-\cos \Omega _{2}t)\approx \frac{1}{2}%
(1-\cos \frac{\Omega _{m}^{2}}{2\Omega _{c}}t).  \label{twophoton}
\end{equation}%
The oscillatory behavior is quite normal (i.e., sinusoidal) except the
square dependence of the Rabi frequency on the microwave amplitude, which is
the signature of two-photon transitions. Therefore, when the microwave field
is weak, we expect two-photon Rabi oscillations to occur, as reported in
\cite{PhysRevB.80.172506}. Furthermore, Eq. (7) predicts that the stronger
the qubit-TLS coupling strength is (larger $\Omega _{c}$), the smaller the
oscillation frequency becomes (Fig. 2(e)). Notice that both of these
results are counterintuitive because one usually would expect that a
stronger qubit-TLS coupling would lead to a faster oscillation and that
two-photon transitions would be significant in strong ac fields.
These intriguing phenomena can be understood with the help of the energy
level structure shown in the lower inset of Fig. 3. When a microwave with $%
2\omega _{q}$ matching the energy difference between $\left\vert
0g\right\rangle $ and $\left\vert 1e\right\rangle $ is applied, population
can be transferred from $\left\vert 0g\right\rangle $ to $\left\vert
1e\right\rangle $ with the help of the intermediate states $\frac{1}{\sqrt{2}%
}$($\left\vert 1g\right\rangle \pm \left\vert 0e\right\rangle $), although
there is no direct coupling between $\left\vert 0g\right\rangle $ and $%
\left\vert 1e\right\rangle $. However, both of the two intermediate states
are off-resonant with the microwave field. Thus the larger $\Omega
_{c}$, the greater the detuning, and the smaller the frequency of
the two-photon Rabi oscillations \cite{PhysRevA.54.4854}. Since the
decoherence time of our qubit is relatively short, we use the spectroscopy
data to demonstrate our prediction. The spectroscopy data in Fig. 3
were obtained using long microwave pulses. The stationary population
generated by the microwave induced transitions is measured. Notice
that in the range of frequencies measured two splittings resulting
from qubit-TLS coupling can be clearly observed \cite{Nat.Commun.}, with $%
2\Omega _{c}/2\pi $ about 20 MHz at 16.590 GHz and 64 MHz at 16.510 GHz,
respectively. Inside each splitting there is a stripe in the avoided
crossing. These stripes are resonant peaks generated by the two-photon
transitions because the positions of the peaks (in frequency) match exactly
to one half of the energy difference between $\left\vert 0g\right\rangle $
and $\left\vert 1e\right\rangle $ in the entire flux bias range.
The stripe inside the 20 MHz splitting has a higher intensity than that
inside the 64 MHz splitting, which agrees with our prediction as discussed
above. We calculated the stationary population of the driven qubit-TLS
system using the Markovian master equation. Shown in the upper inset of Fig.
3 is the dependence of the spectrum intensity on $\Omega _{c}$. The peak
induced by the two-photon transitions (the middle one) disappears as $\Omega
_{c}$ increases. On the other hand, the two-photon transition gradually
vanishes while the single-photon transition becomes dominant as the
microwave power increases. Further increasing microwave power leads to the
anomalous Rabi oscillations and Rabi beating.

In summary, we present a theoretical model to describe the quantum
dynamics of a resonantly driven superconducting qubit coupled to a TLS. The
analytical result gives a clear physical picture of the
system's dynamical behavior and predicts that in a four-level bipartite quantum
system depending on the relative strength of the resonant ac driving and the
interparticle coupling, high-contrast Rabi beating, anomalous Rabi
oscillations, and extraordinary two-photon transitions can occur.
All of these phenomena have been unambiguously observed in our experiment
using a superconducting phase qubit coupled to the TLSs and the data agree
remarkably well with the quantitative prediction of the model. We emphasize
that our model not only provides a unified theoretical description of and physical insights into various experimental observations in the
superconducting qubits reported in the literature, but can also be applied
to understand the dynamics of many other resonantly driven
four-level quantum bipartite systems. The model thus forms a solid theoretical
foundation and provides clear physical intuition to the design and analysis
of coupled bipartite qubit systems for the quantum information processing.

\begin{acknowledgments}
This work is partially supported by NCET, NSFC (10704034, 10725415),
the State Key Program for Basic Research of China (2006CB921801) and NSF Grant No. DMR-0325551.
We acknowledge Northrop Grumman ES in Baltimore MD for technical and foundry support and thank R. Lewis, A. Pesetski, E. Folk, and J. Talvacchio for technical assistance.
\end{acknowledgments}


\clearpage
\begin{figure}
\includegraphics[width=3.4in]{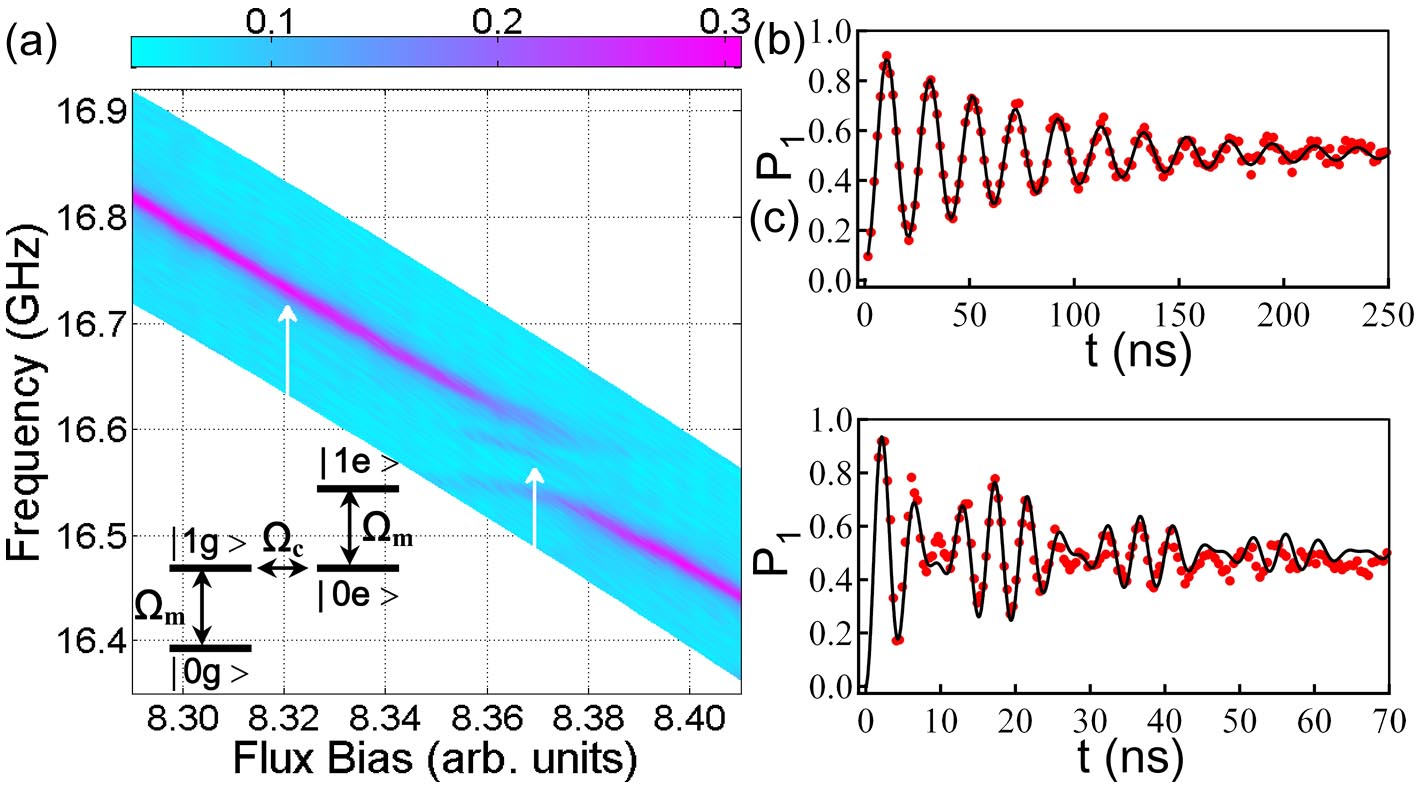}
\caption{\label{fig:epsart} Spectroscopy and coherent oscillations.
(a) Spectroscopy of the qubit versus the flux bias with a splitting at $f=16.572 $ GHz$ $ due to the coupling of the qubit-TLS. (b) Usual Rabi oscillation with the damping time $T_R\approx81.5 $ ns$ $ at $f=16.728 $ GHz$ $ (arrow in the left) where the effect of the TLS is negligible. (c) Coherent oscillation at the avoided crossing (arrow in the right) shows quantum beating due to the interference of Rabi oscillations in the coupled system. In (b) and (c), the red dots  are the experimental results and the solid lines are the theoretical results.}
\end{figure}

\begin{figure}
\includegraphics[width=3.4in]{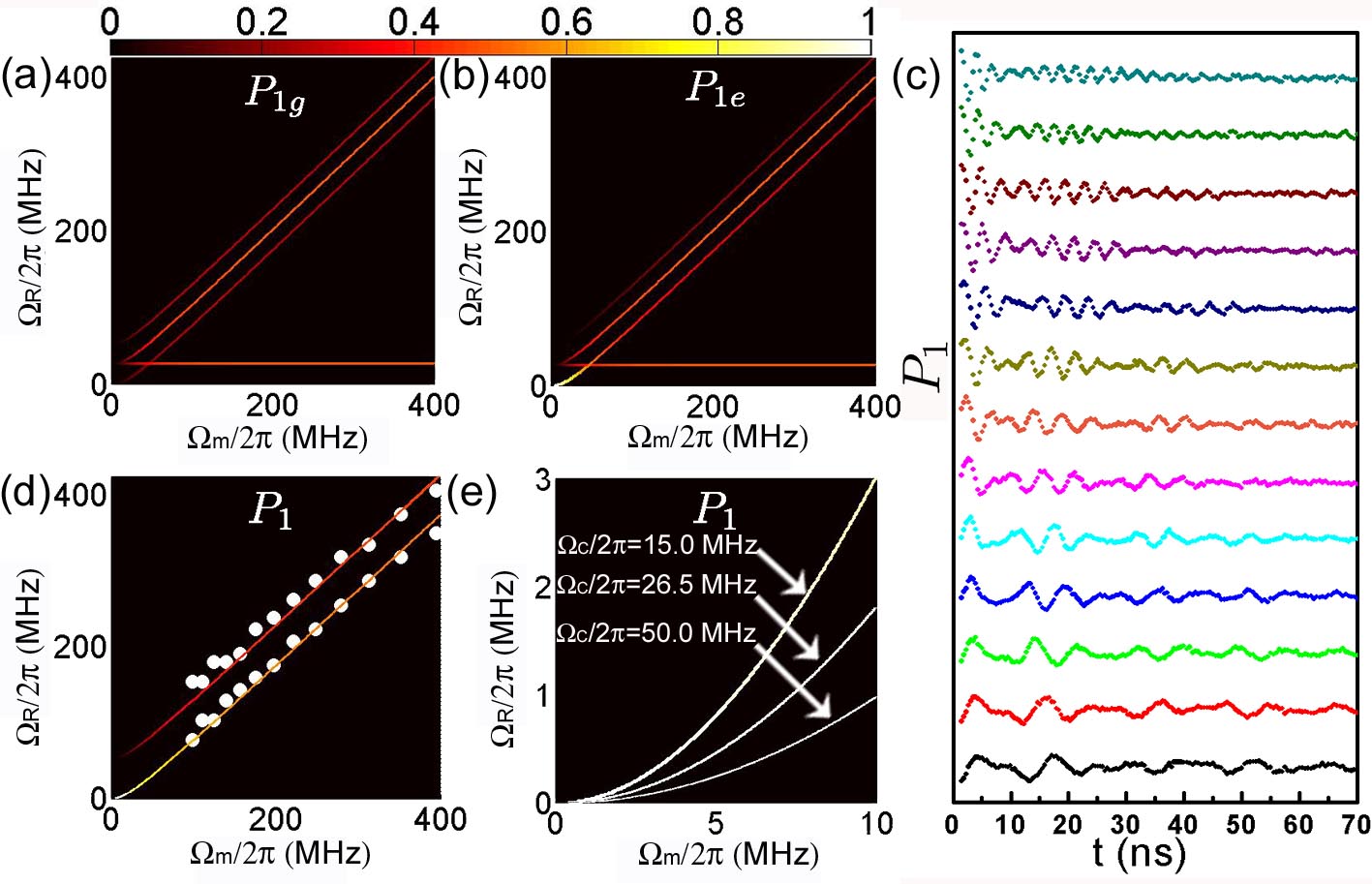}
\caption{\label{fig:epsart}Frequencies $\Omega_R$ in $P_{\phi }$.
(a) and (b), Four frequencies with different weight (indicated by the color) in $P_{1g}$ and $P_{1e}$ versus $\Omega_m$, respectively. (c) Rabi oscillations with the microwave power, at the top of the fridge,
increasing from -13 dBm to -1 dBm with a step of 1 dBm from bottom to top. Curves are shifted vertically for clarity. Quantum beating becomes more clear as the  amplitude of microwave increases. (d) Frequencies (dots), obtained by the Fourier transformations of the corresponding Rabi oscillations in (c), versus the microwave amplitude. The color lines are the two frequencies in $P_1$ obtained from the theoretical analysis. (e) Frequencies in $P_1$ induced by the two-photon transitions versus $\Omega_m$ with three different $\Omega_c/2\pi$: 15.0 MHz, 26.5 MHz, and 50.0 MHz. }
\end{figure}

\begin{figure}
\includegraphics[width=3.4in]{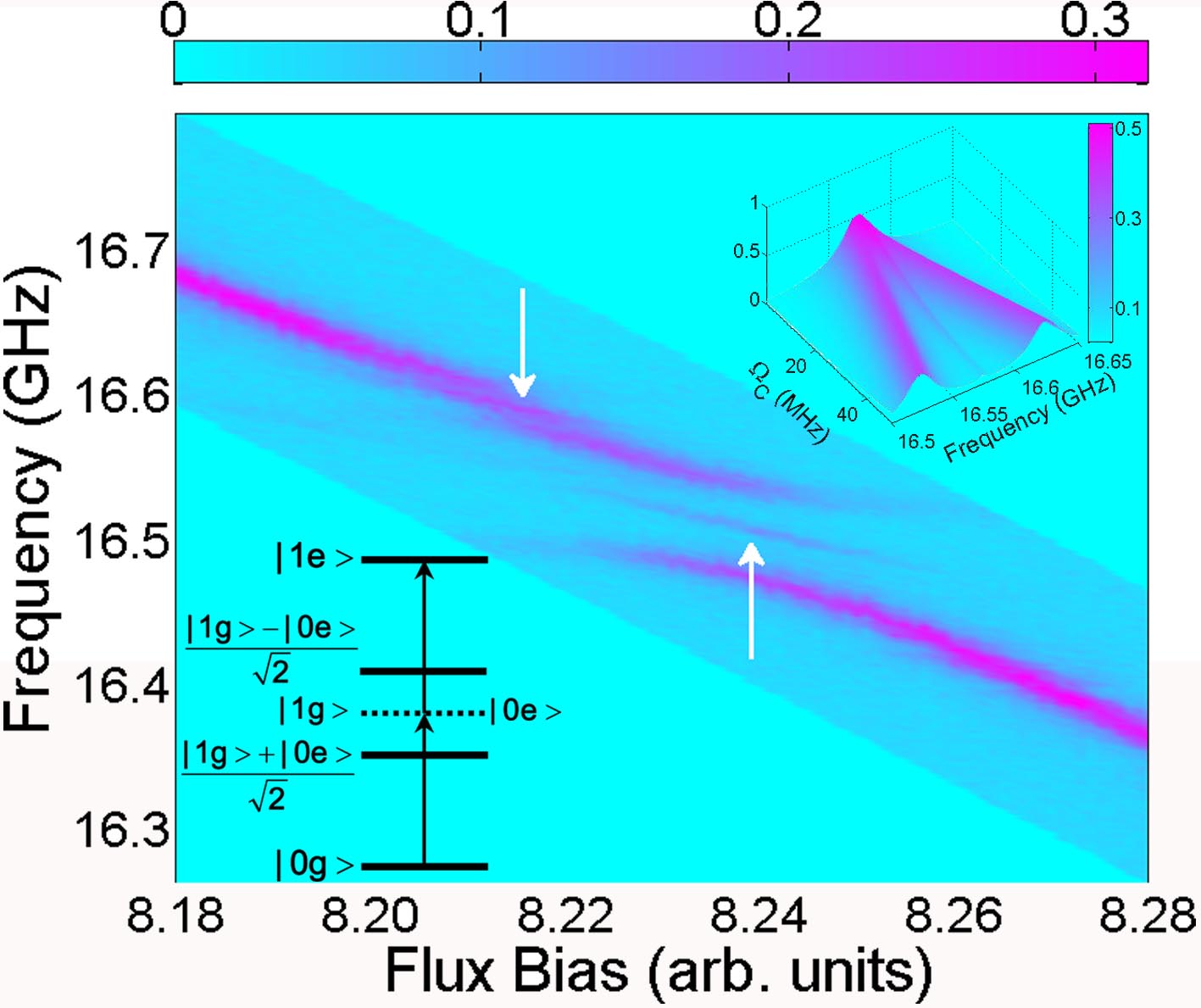}
\caption{\label{fig:epsart}Two-photon transitions.
Large $\Omega_c$ hinders two-photon transitions, leading to a lower resonant peak in the measured spectrum as marked with arrows. The lower inset shows the energy levels in the qubit-TLS coupled system. The upper inset shows the spectrum density versus $\Omega_c$. The outside peaks are due to the stationary population from $\left\vert 0g\right\rangle$ to $\frac{1}{\sqrt{2}}$($\left\vert 1g\right\rangle \pm \left\vert 0e\right\rangle $). The middle peak is due to the two-photon transitions from $\left\vert 0g\right\rangle$ to $\left\vert 1e\right\rangle$.}
\end{figure}

\end{document}